\documentclass[11pt]{article}

\usepackage[T1]{fontenc}
\usepackage[utf8]{inputenc}
\usepackage{lmodern}
\usepackage[a4paper,margin=1in]{geometry}
\usepackage{amsmath,amssymb,bm}
\usepackage{graphicx}
\usepackage{booktabs}
\usepackage{makecell}
\usepackage{rotating}
\usepackage[round,authoryear]{natbib}
\usepackage{url}
\usepackage[hidelinks]{hyperref}
\usepackage{setspace}
\usepackage{microtype}

\title{Clicks, Carbon, and Risk: Climate Attention, Persistent Financial Risk, and Real-Sector Activity}
\author{Giovanni Angelini\thanks{Department of Economics, University of Bologna. Email: g.angelini@unibo.it} \and Luca De Angelis\thanks{Department of Economics, University of Bologna. Email: l.deangelis@unibo.it}}
\date{}

\newcommand{\estse}[2]{%
\begin{tabular}[c]{@{}c@{}}%
{\normalsize #1}\\[-1pt]%
{\scriptsize (#2)}%
\end{tabular}}

\begin{document}
\maketitle

\begin{abstract}
Do shifts in public climate concern alter persistent financial risk and subsequently matter for real activity? We combine daily U.S. sectoral equity returns with a monthly Climate Concern Index (CCI) constructed from internet search intensity and decomposed into adverse and transition-oriented components. GARCH--MIDAS estimates reveal substantial heterogeneity: adverse climate concern raises persistent volatility in carbon-intensive Energy activities, while transition-oriented attention tends to reduce long-run volatility in regulated Utilities, with important differences across sub-sectors. We then construct a persistent-risk differential between Energy and Utilities and examine its predictive content for sectoral real activity. Innovations associated with adverse climate concern precede declines in both Oil and Gas industrial production and employment over the following quarters, while comparable responses are not observed for Utilities. The production result is preserved when the financial predictor is reconstructed from the more closely matched Oil, Gas, and Coal equity sub-sector. Overall, the findings establish a link between climate concern, persistent sectoral financial risk, and subsequent real-sector activity.
\end{abstract}


\section{Introduction}

Climate-related risks affect economic activity through both physical damages and the transition toward a low-carbon economy. Their effects are also transmitted through financial markets, where changing expectations about regulation, technology, future cash flows, and physical exposure can alter asset prices, risk premia, and financing conditions. A growing literature documents that carbon exposure and climate-related news are priced in equity and option markets and can materially affect the risk borne by carbon-intensive firms and sectors \citep{BoltonKacperczyk2021,MonasteroloDeAngelis2020,IlhanSautnerVilkov2021,PastorStambaughTaylor2021}.

A related strand of research studies climate attention and climate concern. Measures constructed from news, text, and internet-search activity capture changes in the salience of climate risks and in investors' perceptions of those risks \citep{EngleEtAl2020,ArdiaEtAl2023,DaEngelbergGao2011}. This distinction matters because perceptions may change even in the absence of a contemporaneous aggregate physical shock and may therefore represent an independent source of financial and macroeconomic uncertainty. In earlier work, \citet{AngeliniEtAl2025} construct the Climate Concern Index (CCI) from disaggregated Google searches and show that exogenous increases in climate concern are followed by lower employment and private consumption together with a persistent increase in stock-market volatility. The present paper complements that aggregate evidence by asking where, within the equity market, this increase in persistent risk is concentrated and whether the resulting sectoral risk signal contains information about subsequent real activity.

Our focus is on the long-run component of sectoral equity volatility. We use U.S. Energy and Utilities indices together with a set of more granular sub-sector indices that differ markedly in their exposure to carbon-intensive activity, regulated infrastructure, transition investment, and physical climate risk. We further decompose the CCI into a negative component (CCI$^-$), capturing attention to physical risks, natural disasters, and adverse climate narratives, and a positive component (CCI$^+$), capturing attention to mitigation, emissions reduction, and technological solutions. This decomposition allows us to examine whether adverse climate concern and transition-oriented attention have different effects across sectors with different exposures to physical and transition risks.

Methodologically, we estimate GARCH--MIDAS models that combine daily sectoral returns with monthly climate-attention measures. The framework separates transitory high-frequency volatility from a slow-moving component linked to climate concern. This is particularly useful in the present setting because changes in expected regulation, transition costs, investment opportunities, and physical exposure may affect valuations over horizons much longer than a few trading days.

This paper contributes in three ways. First, we isolate the persistent component of sectoral equity volatility associated with climate concern, rather than focusing on aggregate or short-run market volatility. Second, we document substantial heterogeneity both across and within Energy and Utilities, showing that the sign and magnitude of the persistent-volatility response depend on the type of climate concern and on sectoral exposure to physical and transition risks. Third, we connect the financial evidence to real-sector outcomes by examining whether innovations in the persistent financial-risk differential between Energy and Utilities contain predictive information for subsequent sectoral activity.

The financial results reveal a pronounced asymmetry. For aggregate Energy, the long-run volatility response is positive for all CCI variants and is strongest for CCI$^-$. For aggregate Utilities, by contrast, the full CCI and CCI$^+$ are associated with lower persistent volatility, while the coefficient on CCI$^-$ is not statistically significant. The sub-sector evidence sharpens this contrast: upstream and integrated Energy activities are particularly sensitive to adverse climate concern, Pipelines behave more like regulated infrastructure, and Water displays a positive response consistent with direct physical-risk exposure. This cross-sector contrast motivates the persistent-risk differential used in the real-activity analysis.

The real-activity evidence reinforces this sectoral pattern. Innovations in the adverse-climate persistent-risk differential precede declines in both Oil and Gas industrial production and employment over the following quarters, whereas comparable responses are not observed for Utilities. The production result is also preserved when the financial predictor is reconstructed using the more closely matched Oil, Gas, and Coal equity sub-sector. Moreover, the persistent-risk differential retains predictive content after controlling for the underlying CCI itself, suggesting that the financial transformation of climate concern contains information about subsequent sectoral activity beyond the original attention measure.

These findings are consistent with a macro-financial link between climate concern, persistent sectoral financial risk, and real activity. A rise in persistent financial risk may be associated with higher required returns, tighter financing conditions, or delayed investment, although these intermediate mechanisms are not separately identified here. We therefore interpret the real-activity evidence as predictive rather than as establishing a causal mediation mechanism. This interpretation connects our sectoral financial evidence to the broader uncertainty literature, in which elevated uncertainty can depress real activity \citep{Bloom2009}, and to recent work showing that climate-policy uncertainty can operate through financial conditions and investment in carbon-intensive sectors \citep{Mallea2026}.

The remainder of the paper is organized as follows. Section~\ref{cci} describes the CCI and its decomposition. Section~\ref{midas} presents the GARCH--MIDAS framework. Section~\ref{res} reports the sectoral and sub-sectoral financial results. Section~\ref{macro} examines whether persistent climate-related financial risk contains information for subsequent sectoral real activity. Section~\ref{concl} concludes.

\section{Climate Concern Index (CCI)}
\label{cci}

This section describes the construction of the Climate Concern Index (CCI), a monthly measure of public climate-related concern based on Google search activity, proposed by \citet{AngeliniEtAl2025}. The CCI builds on the methodology of \citet{bontempi2021eurq} and interprets search volumes as signals of \emph{epistemic uncertainty}, that is, situations in which individuals seek information about potentially knowable but unsettled issues that may affect them. In this sense, web searches capture spontaneous and widespread reactions to climate-related events and narratives, providing a dynamic proxy for the salience of climate-related concerns, risks, and opportunities.

A crucial step in constructing the CCI is the development and validation of an appropriate vocabulary. \citet{AngeliniEtAl2025} implement a multi-stage procedure. First, the existing literature on climate sentiment and attention measures is reviewed to compile a preliminary list of keywords. Second, this list is refined and expanded using prompt-engineering techniques with large language models to identify emerging and contextually relevant terminology. Third, the empirical salience of the candidate terms is validated by examining their effective search frequency and relevance in online discourse. This iterative procedure leads to a final set of 112 queries for the United States, grouped into seven thematic categories:
\begin{enumerate}
\item[(i)] Natural disasters (e.g., extreme weather, floods);
\item[(ii)] Global warming threats (e.g., rising seas, deforestation);
\item[(iii)] General climate change narratives (e.g., climate crisis, climate refugees);
\item[(iv)] Mitigation and reduction strategies (e.g., climate solutions, circular economy);
\item[(v)] Technological hope (e.g., biodiversity, aquaculture);
\item[(vi)] International summits (e.g., Paris Agreement, Kyoto Protocol);
\item[(vii)] Climate and environmental policy (e.g., decarbonization, IPCC).
\end{enumerate}

In the baseline specification, all seven dimensions are aggregated to construct the overall CCI.\footnote{Further details on the construction and validation of the CCI are provided in \citet{AngeliniEtAl2025}.} In this paper, however, we exploit the thematic classification to construct two economically meaningful sub-indices. Specifically, the \emph{Negative CCI} (CCI$^-$) is based on categories (i)--(iii), which capture attention to physical risks, natural disasters, and adverse climate narratives. The \emph{Positive CCI} (CCI$^+$) is constructed from categories (iv) and (v), which reflect attention to mitigation strategies, emissions reduction, and technological progress toward a low-carbon economy. The remaining categories, (vi) and (vii), are mainly related to institutional and policy communication and are excluded from the decomposition because they do not map unambiguously into either an adverse or a solution-oriented dimension.

\begin{figure}[!htbp]
    \centering
    \includegraphics[width=\textwidth]{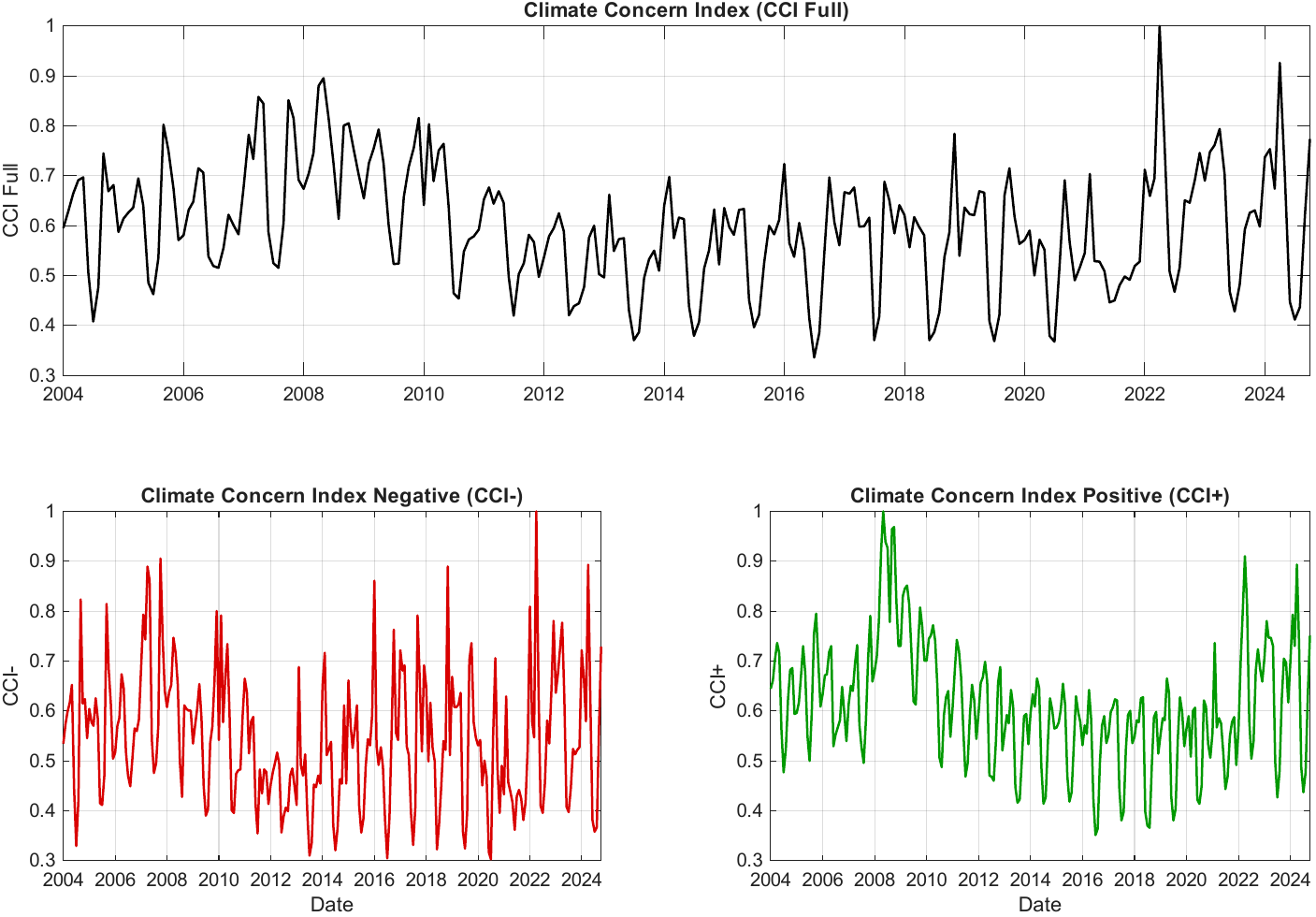}
    \caption{Climate Concern Index (CCI) and its decomposition. The top panel reports the full CCI, while the bottom panels report the negative component (CCI$^-$) and the positive component (CCI$^+$). Monthly frequency, January~2004--October~2024.}
    \label{fig:cci_decomposition}
\end{figure}

Figure~\ref{fig:cci_decomposition} reports the monthly dynamics of the three indices over the period January~2004--October~2024, while Table~\ref{tab:cci} presents their descriptive statistics. The full CCI exhibits three broad phases: increasing concern between 2004 and 2010, relative stagnation from 2011 to 2020, and renewed growth from late 2021 onward. As discussed in \citet{AngeliniEtAl2025}, spikes are often associated with extreme weather events or coordinated environmental campaigns such as Earth Day, consistent with the view that salient climate-related events and communication are associated with stronger information-seeking behavior.

Although both sub-indices broadly track the full CCI, their dynamics differ in important ways. The CCI$^-$ is characterized by sharper and more short-lived spikes, typically associated with extreme weather events and other adverse climate-related episodes. By contrast, the CCI$^+$ displays smoother and more persistent movements, reflecting a more gradual and sustained interest in mitigation strategies and low-carbon technologies. In particular, the negative component reacts more abruptly during episodes of heightened physical-risk salience, whereas the positive component displays a more stable upward pattern in the post-2021 period, consistent with growing attention to transition policies and innovation.

The correlations reported in the bottom panel of Table~\ref{tab:cci} show that both CCI$^-$ and CCI$^+$ are strongly correlated with the full CCI, with coefficients of 0.92 and 0.87, respectively. Their mutual correlation is substantially lower, at 0.62, suggesting that the two components capture related but distinct dimensions of climate-related search behavior. The CCI$^-$ also displays slightly greater variability and more pronounced positive skewness, consistent with the episodic nature of adverse climate-related attention. Since each index is normalized independently to have a maximum value equal to one, differences in average levels should not be interpreted as differences in absolute intensity across components. Overall, the decomposition reveals distinct dynamics in adverse and solution-oriented climate concern, motivating their separate inclusion in the empirical analysis and allowing us to examine potentially asymmetric effects across sectors with different exposures to physical and transition risks.

\begin{table}[!htbp]
    \centering
    \begin{tabular}{lccccccc}
    \hline
    Index & Mean & Std.Dev. & Min & Skewness & Kurtosis & 5\% Perc. & 95\% Perc. \\
    \hline
    CCI Full   & 0.5997 & 0.1197 & 0.3359 & 0.2043 & 2.937 & 0.4022 & 0.8027 \\
    CCI$^-$    & 0.5451 & 0.1314 & 0.3030 & 0.5726 & 3.202 & 0.3565 & 0.7923 \\
    CCI$^+$    & 0.6123 & 0.1208 & 0.3515 & 0.4309 & 3.435 & 0.4175 & 0.8176 \\
    \hline
    \multicolumn{4}{c}{Correlations:} & & & \\
    \hline
               & CCI Full & CCI$^-$ & CCI$^+$ & & & \\
    CCI Full   & 1.0000   & 0.9151  & 0.8709  & & & \\
    CCI$^-$    &          & 1.0000  & 0.6200  & & & \\
    CCI$^+$    &          &         & 1.0000  & & & \\
    \hline
    \end{tabular}
    \caption{Descriptive statistics for the full CCI and the two sub-indices CCI$^-$ and CCI$^+$. Monthly frequency, January~2004--October~2024.}
    \label{tab:cci}
\end{table}

\section{GARCH--MIDAS framework}
\label{midas}

We model sectoral volatility using the GARCH--MIDAS framework of
\citet{EngleGhyselsSohn2013}, which separates high-frequency volatility
dynamics from a slow-moving component driven by lower-frequency variables.

Let $R_{it}$ denote the observed daily return of sector $i$ on trading day $t$. Before estimating the volatility model, we remove the component of sectoral returns explained by the five Fama--French factors. Denoting the vector of daily factors by $F_t$ and the corresponding estimated factor loadings by $\widehat{\boldsymbol{\beta}}_i$, we define the factor-adjusted return as
\[
r_{it}
=
R_{it}
-
F_t^{\prime}\widehat{\boldsymbol{\beta}}_i.
\]
The GARCH--MIDAS model is then estimated on the factor-adjusted return $r_{it}$:
\[
r_{it}
=
\mu_i+\varepsilon_{it},
\qquad
\varepsilon_{it}
=
\sqrt{h_{it}}\,z_{it},
\qquad
\mathbb{E}(z_{it}\mid\mathcal F_{t-1})=0,
\qquad
\operatorname{Var}(z_{it}\mid\mathcal F_{t-1})=1.
\]

The conditional variance is decomposed into a high-frequency component and a slow-moving component,
\[
h_{it}
=
\tau_{i,m(t)}^{\,j}g_{it},
\]
where $m(t)$ associates trading day $t$ with the corresponding monthly period and
$j\in\{\mathrm{Full},-,+\}$ indexes the CCI specification. In the empirical implementation, monthly periods are approximated by blocks of 20 trading days, consistently with the mixed-frequency alignment of daily returns and the monthly CCI. The component
$\tau_{i,m}^{\,j}$ captures persistent movements in sectoral variance, while $g_{it}$ captures short-run daily volatility dynamics.

Let $X_m^{\,j}$ denote the corresponding monthly climate-attention measure. We construct the MIDAS distributed lag
\[
\widetilde X_m^{\,j}
=
\sum_{k=1}^{K}
w(k;\omega_i^{\,j})X_{m-k}^{\,j},
\qquad
w(k;\omega_i^{\,j})\geq0,
\qquad
\sum_{k=1}^{K}
w(k;\omega_i^{\,j})=1,
\]
where $k=1$ denotes the most recent lag. Consistently with the empirical implementation, the weights follow a one-parameter Beta polynomial,
\[
w(k;\omega_i^{\,j})
=
\frac{
\left(
1-\frac{k}{K}+\delta
\right)^{\omega_i^{\,j}-1}
}{
\displaystyle
\sum_{\ell=1}^{K}
\left(
1-\frac{\ell}{K}+\delta
\right)^{\omega_i^{\,j}-1}
},
\qquad
k=1,\ldots,K,
\]
where $\delta>0$ is a small constant introduced for numerical stability. We set $K=48$, so that the lag window spans the previous four years of climate attention, with the estimated Beta polynomial determining how weight is distributed across that window.

The persistent variance component is specified using a log link,
\[
\tau_{i,m}^{\,j}
=
\exp\left(
\kappa_i^{\,j}
+
\theta_i^{\,j}\widetilde X_m^{\,j}
\right).
\]
The coefficient $\theta_i^{\,j}$ is the main parameter of interest. A positive value implies that a higher level of the corresponding CCI measure is associated with higher persistent sectoral variance, whereas a negative value implies a lower persistent variance component. The parameter $\omega_i^{\,j}$ determines the shape of the MIDAS weighting scheme and hence how strongly more recent observations of climate attention are weighted relative to more distant ones.

The short-run component follows a scaled GARCH(1,1) recursion,
\[
g_{it}
=
(1-\alpha_i^{\,j}-\beta_i^{\,j})
+
\frac{\alpha_i^{\,j}}
{\tau_{i,m(t)}^{\,j}}
\varepsilon_{i,t-1}^{2}
+
\beta_i^{\,j}g_{i,t-1},
\]
with
\[
\alpha_i^{\,j}\geq0,
\qquad
\beta_i^{\,j}\geq0,
\qquad
\alpha_i^{\,j}+\beta_i^{\,j}<1.
\]
The intercept normalization implies that the short-run component fluctuates around one, leaving $\tau_{i,m}^{\,j}$ to determine the persistent scale of conditional variance.

For each sector, the model is estimated separately for the Full, Negative, and Positive CCI specifications by maximum likelihood under conditional Gaussianity. The resulting estimates
$\widehat{\tau}_{i,m}^{\,j}$ provide the sector-specific persistent-risk components used in the real-activity analysis of Section~\ref{macro}.

\section{Empirical results}
\label{res}

We study U.S. sectoral equity returns over the period January~2004--October~2024.
Daily sector price indices are obtained from Datastream. The baseline analysis focuses
on the U.S. Datastream Energy and Utilities indices.\footnote{The Datastream mnemonics
are \texttt{ENEGYUS} for US--DS Energy and \texttt{UTILSUS} for US--DS Utilities.
Both series are price indices denominated in U.S. dollars.}
Daily returns are computed as 100 times log differences of the corresponding price indices.
As described in Section~\ref{midas}, the GARCH--MIDAS models are estimated on returns
adjusted for the Fama--French five factors.\footnote{The daily Fama--French factors are obtained from Kenneth French's Data Library.}

\begin{figure}[!htbp]
    \centering
    \includegraphics[width=\textwidth]{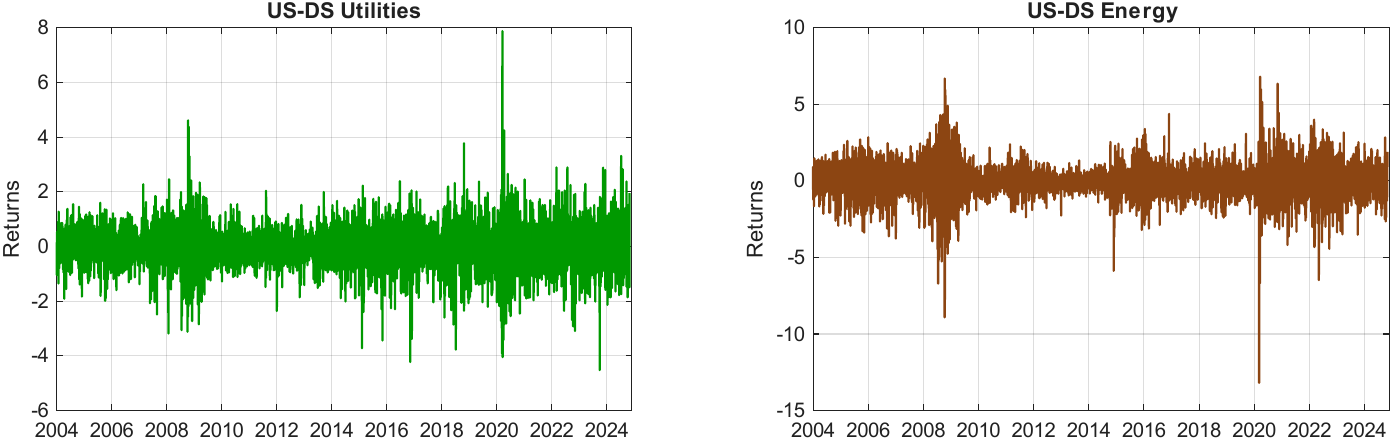}
    \caption{US--DS Utilities and US--DS Energy daily returns and the monthly Climate Concern Index (CCI).}
    \label{fig:returns_cci}
\end{figure}

The mixed-frequency specification allows monthly climate attention to enter the slow-moving variance component, while short-run return innovations are captured by the daily GARCH dynamics. For each sector, we estimate three GARCH--MIDAS specifications that differ
only in the monthly climate-attention measure entering the persistent
variance component: the full CCI, CCI$^-$, and CCI$^+$. 
Table~\ref{tab:aggregate_midas} reports the estimated MIDAS slope coefficient
$\theta$ for the two broad sectors and the three CCI specifications.
Consistent with the discussion in Section~\ref{midas}, $\theta$ is the main
parameter of interest, as it determines how each climate-attention measure is
associated with the persistent component of sectoral variance. Full
GARCH--MIDAS parameter estimates are reported in Appendix
Table~\ref{tab:aggregate_midas_full}.

For Utilities, $\theta$ is negative and statistically significant under the full CCI ($\theta=-6.5077$) and CCI$^+$ ($\theta=-5.6869$), while the coefficient associated with CCI$^-$ is positive but statistically insignificant ($\theta=0.5803$). Thus, persistent Utilities risk declines when broad climate attention or solution-oriented climate attention increases, whereas adverse climate concern has no precisely estimated effect at the aggregate Utilities level. This pattern is consistent with the relatively regulated nature of the sector and with the possibility that greater attention to mitigation and low-carbon technologies improves the visibility of transition-related investment opportunities and future cash flows.

Energy displays a markedly different pattern. The coefficient $\theta$ is positive and statistically significant for all three CCI measures. It equals $1.3343$ for the full index, rises to $4.8184$ for CCI$^-$, and remains positive at $1.1141$ for CCI$^+$. The substantially larger coefficient associated with CCI$^-$ indicates that adverse climate narratives and physical-risk salience are particularly informative for the persistent risk of carbon-intensive activities. At the same time, the positive response to CCI$^+$ shows that solution-oriented climate attention need not reduce uncertainty for the Energy sector. Greater attention to mitigation and technological progress may instead heighten uncertainty about the pace of substitution away from fossil fuels, future regulation, and the reallocation of investment toward low-carbon technologies.

Taken together, the aggregate results reveal a clear sectoral asymmetry. Climate attention is associated with higher persistent risk in Energy, especially when it reflects adverse climate concern, whereas broad and solution-oriented climate attention are associated with lower persistent risk in Utilities. The sub-sector analysis below examines whether this contrast reflects systematic differences in business models and exposure to physical and transition risks within the two broad sectors.

\begin{table}[!htbp]
\centering
\caption{Climate concern and persistent volatility: aggregate sector estimates.}
\label{tab:aggregate_midas}
\renewcommand{\arraystretch}{1.15}
\setlength{\tabcolsep}{14pt}

\begin{tabular}{lccc}
\toprule
Sector & Full CCI & CCI$^-$ & CCI$^+$ \\
\midrule

Utilities
& \estse{-6.5077}{1.3172}
& \estse{0.5803}{0.3700}
& \estse{-5.6869}{0.8704} \\

Energy
& \estse{1.3343}{0.3788}
& \estse{4.8184}{1.6667}
& \estse{1.1141}{0.3763} \\

\bottomrule
\end{tabular}

\begin{minipage}{0.88\textwidth}
\footnotesize
\emph{Notes:} The table reports estimates of the MIDAS slope coefficient
$\theta$, which measures the association between each climate-attention
measure and the persistent component of sectoral variance.
Standard errors are reported in parentheses.
Full GARCH--MIDAS parameter estimates are reported in
Appendix Table~\ref{tab:aggregate_midas_full}.
\end{minipage}
\end{table}

\subsection{Sub-sector heterogeneity}
\label{sub}

\begin{figure}[!htbp]
\centering
\includegraphics[width=\textwidth]{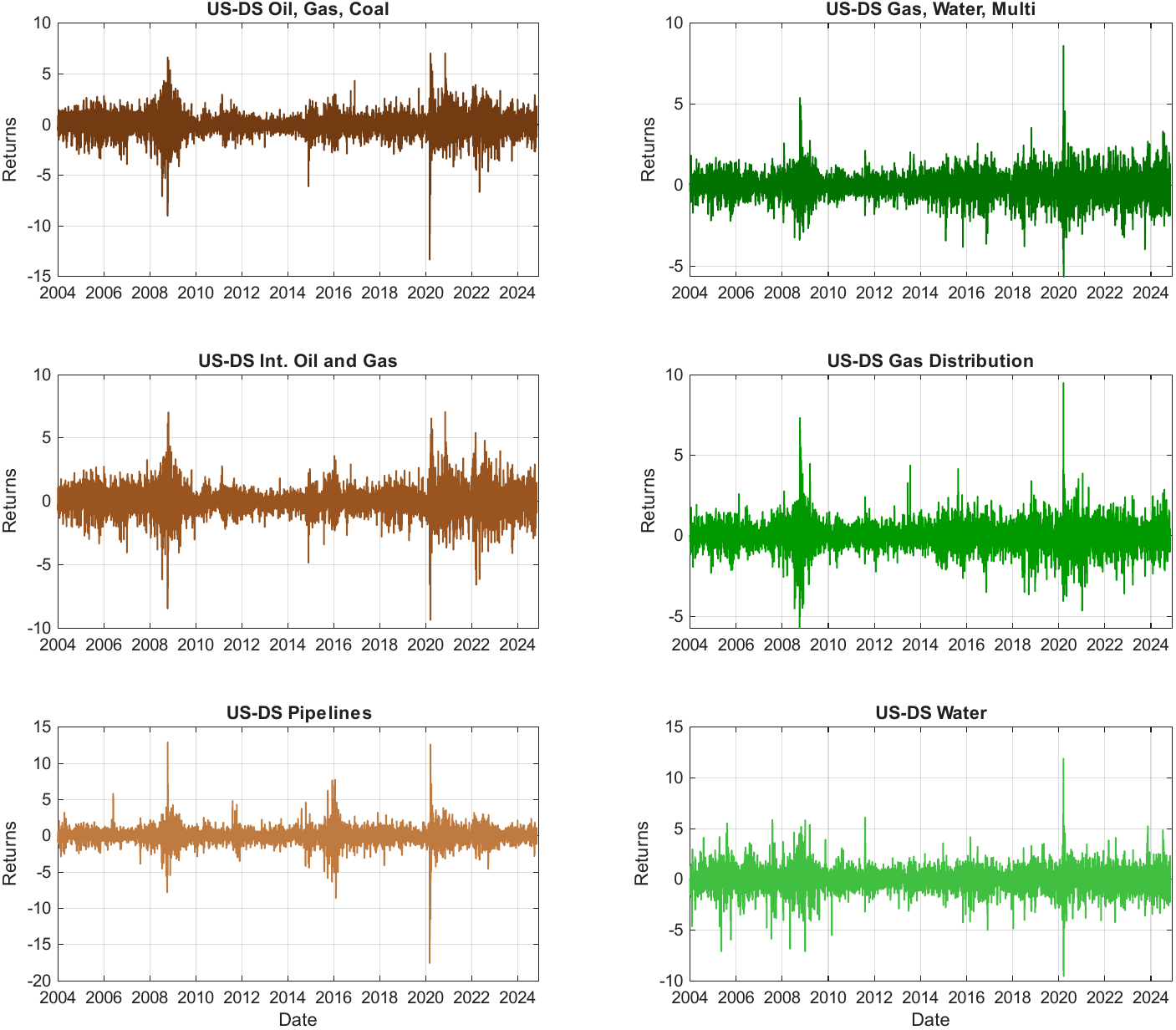}
\caption{Daily log-returns for Energy (left column) and Utilities (right column) sub-sector indices, January~2004--October~2024. Each panel reports one sub-sector return series.}
\label{fig:returns_6subsectors_3x2}
\end{figure}

The aggregate Energy--Utilities contrast masks substantial heterogeneity within both sectors. We therefore estimate the same GARCH--MIDAS specifications for three Energy sub-sectors (Oil, Gas and Coal; Integrated Oil and Gas; and Pipelines) and three Utilities sub-sectors (Gas, Water and Multi-Utilities; Gas Distribution; and Water).\footnote{The Datastream mnemonics are \texttt{OILGCUS}, \texttt{OILINUS}, and \texttt{PIPELUS} for the Energy sub-sectors, and \texttt{GWMUTUS}, \texttt{GASDSUS}, and \texttt{WATERUS} for the Utilities sub-sectors.} Figure~\ref{fig:returns_6subsectors_3x2} reports the corresponding daily returns, while Table~\ref{tab:univariate_6assets_cci} reports the estimated MIDAS slope coefficient $\theta$ across sub-sectors and CCI specifications. The complete GARCH--MIDAS estimates are reported in Appendix Table~\ref{tab:univariate_6assets_cci_full}.

Within Energy, Oil, Gas and Coal and Integrated Oil and Gas closely reproduce the aggregate-sector pattern. Their estimated responses to the full CCI are positive, while the association becomes considerably stronger for CCI$^-$, with $\theta$ close to 4.4 and 4.9, respectively. Adverse climate concern therefore appears particularly relevant for the persistent risk of the most carbon-intensive Energy activities.

The response to solution-oriented attention is more heterogeneous. Under CCI$^+$, the coefficient for Oil, Gas and Coal becomes strongly negative ($\theta=-6.6716$), whereas it remains positive for Integrated Oil and Gas ($\theta=0.8084$). Pipelines depart even more clearly from the aggregate Energy pattern: their estimated coefficient is negative under all three CCI specifications. These differences suggest that the effect of climate attention within Energy depends materially on the underlying business model and exposure to the transition, rather than simply on broad sector classification.

A similarly sharp contrast emerges within Utilities. Gas, Water and Multi-Utilities and Gas Distribution display negative coefficients across all three climate-attention measures, broadly reinforcing the aggregate Utilities results. Water, however, moves in the opposite direction: its estimated coefficient is positive under the full CCI, CCI$^-$, and CCI$^+$, and is largest for the negative component ($\theta=4.5796$). This pattern is consistent with the direct exposure of water-related activities to physical climate risks, including droughts, floods, and infrastructure stress, which may dominate the stabilizing mechanisms relevant for other regulated Utilities activities.

Overall, the sub-sector evidence qualifies the broad Energy--Utilities distinction in economically meaningful ways. Adverse climate concern is associated with higher persistent risk in upstream and integrated Energy activities, whereas several Utilities segments exhibit the opposite relationship. At the same time, Pipelines depart from the rest of Energy and Water from the rest of Utilities. The response of persistent sectoral risk to climate attention therefore appears to depend not only on sector membership, but also on the specific exposure to physical and transition risks.

\begin{table}[!htbp]
\centering
\caption{Climate concern and persistent sectoral volatility: sub-sector estimates.}
\label{tab:univariate_6assets_cci}
\renewcommand{\arraystretch}{1.15}
\setlength{\tabcolsep}{12pt}

\begin{tabular}{lccc}
\toprule
Sub-sector & Full CCI & CCI$^-$ & CCI$^+$ \\
\midrule

\multicolumn{4}{l}{\textit{Energy}} \\[2pt]

Oil, Gas, Coal
& \estse{1.1455}{0.3513}
& \estse{4.3827}{1.7203}
& \estse{-6.6716}{1.6034} \\

Integrated Oil and Gas
& \estse{1.1326}{0.3232}
& \estse{4.9426}{1.5948}
& \estse{0.8084}{0.3402} \\

Pipelines
& \estse{-3.5002}{1.1391}
& \estse{-2.4235}{1.4415}
& \estse{-3.1972}{0.8459} \\

\addlinespace
\multicolumn{4}{l}{\textit{Utilities}} \\[2pt]

Gas, Water, Multi-Utilities
& \estse{-7.9874}{1.5835}
& \estse{-7.5595}{2.3491}
& \estse{-7.0016}{1.0357} \\

Gas Distribution
& \estse{-5.6926}{1.3218}
& \estse{-5.0599}{1.8743}
& \estse{-4.7346}{1.1221} \\

Water
& \estse{3.1749}{0.5017}
& \estse{4.5796}{0.6743}
& \estse{2.0402}{0.3851} \\

\bottomrule
\end{tabular}

\begin{minipage}{0.90\textwidth}
\footnotesize
\emph{Notes:} The table reports estimates of the MIDAS slope coefficient
$\theta$, which measures the association between each climate-attention
measure and the persistent component of sectoral variance.
Standard errors are reported in parentheses.
Full GARCH--MIDAS parameter estimates are reported in
Appendix Table~\ref{tab:univariate_6assets_cci_full}.
\end{minipage}
\end{table}

\section{Persistent climate-related financial risk and real-sector activity}
\label{macro}

The preceding results show that climate concern is associated with markedly different persistent-risk dynamics across Energy and Utilities. We next examine whether this financial signal contains information about subsequent real-sector activity. Specifically, we ask whether innovations in the relative persistent risk of Energy and Utilities precede changes in sectoral production and employment.

\subsection{Empirical design}

Our primary real-activity outcome is quarterly, seasonally adjusted industrial production for Oil and Gas Extraction from the Federal Reserve's G.17 release (FRED series \texttt{IPG211SQ}). Electric and Gas Utilities (\texttt{IPG2211A2SQ}) provide a sectoral comparison. We complement production with seasonally adjusted employment from the Bureau of Labor Statistics Current Employment Statistics program, using Oil and Gas Extraction employment (\texttt{CES1021100001}) and Utilities employment (\texttt{CES4422000001}), both accessed through FRED. The monthly employment series are aggregated to calendar quarters. The Utilities employment series covers the broader Utilities sector and therefore provides a complementary, rather than exact, counterpart to the Electric and Gas Utilities production series. Employment offers an outcome that is conceptually distinct from industrial production and allows us to assess whether the real-sector signal extends beyond a single measure of activity.

Although the underlying industrial-production series and the estimated persistent variance components are available at monthly frequency, we conduct the baseline analysis quarterly. The GARCH--MIDAS component is constructed using a 48-month lag window and is designed to capture slow-moving rather than short-run financial risk. Our empirical question likewise concerns medium-horizon real-sector dynamics rather than month-to-month fluctuations. Quarterly aggregation therefore provides a natural frequency for the baseline exercise and reduces sensitivity to short-run variation in an estimated persistent component. Appendix~\ref{app:freq} examines this choice directly. When the quarterly financial-risk innovation is retained but industrial production is measured at monthly horizons, the estimated response closely tracks the quarterly baseline at comparable horizons. By contrast, innovations extracted directly from the monthly persistent-risk differential generate less stable responses.

To construct the financial predictor, we first aggregate the estimated GARCH--MIDAS persistent components to calendar quarters. For each CCI specification $j\in\{\mathrm{Full},-,+\}$, the persistent-risk differential (PRD) is defined as
\[
PRD_q^j
=
\frac{1}{3}\sum_{m\in q}
\left[
\log \widehat{\tau}^{\,j}_{E,m}
-
\log \widehat{\tau}^{\,j}_{U,m}
\right],
\]
where $E$ and $U$ denote Energy and Utilities, respectively. Throughout this section, $q$ indexes calendar quarters, while $m$ denotes months and $t$ denotes trading days, consistently with the notation introduced in Section~\ref{midas}. A higher $PRD_q^j$ therefore indicates higher persistent financial risk in Energy relative to Utilities.

The PRD series are highly persistent. We therefore fit autoregressive models to each differential and use the standardized residuals as innovations. For each series, we consider AR($p$) specifications with $p=1,\ldots,8$. Lag length is selected on the basis of the Akaike and Bayesian information criteria, while also requiring the residuals to display no significant remaining serial correlation according to Ljung--Box tests at lags 4 and 8. This procedure selects AR(5), AR(6), and AR(5) specifications for the Full, Negative, and Positive PRDs, respectively; the corresponding diagnostics are reported in Appendix Table~\ref{tab:prd_ar_diagnostics}. We refer to the resulting standardized residuals as PRD innovations; they are innovations relative to the predictable dynamics of the estimated differential and are not interpreted as externally identified structural shocks.

We then estimate local projections in the spirit of \citet{Jorda2005}:
\begin{equation}
\log Y_{q+h}-\log Y_q
=
\alpha_h
+\beta_h \eta_q^{j}
+\rho_h\Delta\log Y_q
+\sum_{\ell=0}^{4}\gamma_{\ell,h}\Delta\log OIL_{q-\ell}
+\delta_h\Delta\log IP^{US}_q
+\phi_h FF_q
+u_{q+h},
\label{eq:lp_real}
\end{equation}
for horizons $h=1,\ldots,8$. Here $\eta_q^{j}$ denotes the standardized innovation in the corresponding PRD, $Y_q$ is the sectoral real-activity outcome, $OIL$ is the WTI crude-oil price, $IP^{US}$ is aggregate U.S. industrial production, and $FF$ is the Federal Funds rate. The distributed oil-price controls account for the strong dependence of both Energy-sector financial conditions and real activity on the oil cycle, while aggregate industrial production and the Federal Funds rate capture broader macroeconomic conditions.

Inference is based on heteroskedasticity- and autocorrelation-consistent standard errors with a small-sample correction and a bandwidth that increases with the projection horizon. The local projections are interpreted as predictive evidence: $\beta_h$ measures whether an innovation in the relative persistent financial risk of Energy precedes subsequent changes in sectoral activity, conditional on the included controls. The specification is not intended to identify a structural shock or a causal mediation mechanism.

\subsection{Industrial production}

Figure~\ref{fig:realactivity_ip_emp} (left panel) reports the cumulative response of Oil and Gas industrial production to a one-standard-deviation innovation in the CCI$^-$ persistent-risk differential. The production response is concentrated in the specification associated with adverse climate concern. Following the innovation, Oil and Gas industrial production declines progressively, with the cumulative response reaching about $-1.1\%$ after three quarters and around $-1.5\%$ after four to five quarters. The response is most precisely estimated at the three-quarter horizon and remains economically sizeable thereafter. By contrast, Electric and Gas Utilities production shows no comparable medium-horizon response.

This pattern is specific to the negative component of climate concern. Innovations in the Full and CCI$^+$ persistent-risk differentials do not generate a similarly pronounced contraction in Oil and Gas production. The asymmetry is consistent with the financial-market evidence in Section~\ref{res}, where CCI$^-$ produces the strongest increase in persistent risk for carbon-intensive Energy activities.

\subsection{Employment}

\begin{figure}[!htbp]
\centering
\includegraphics[width=\textwidth]{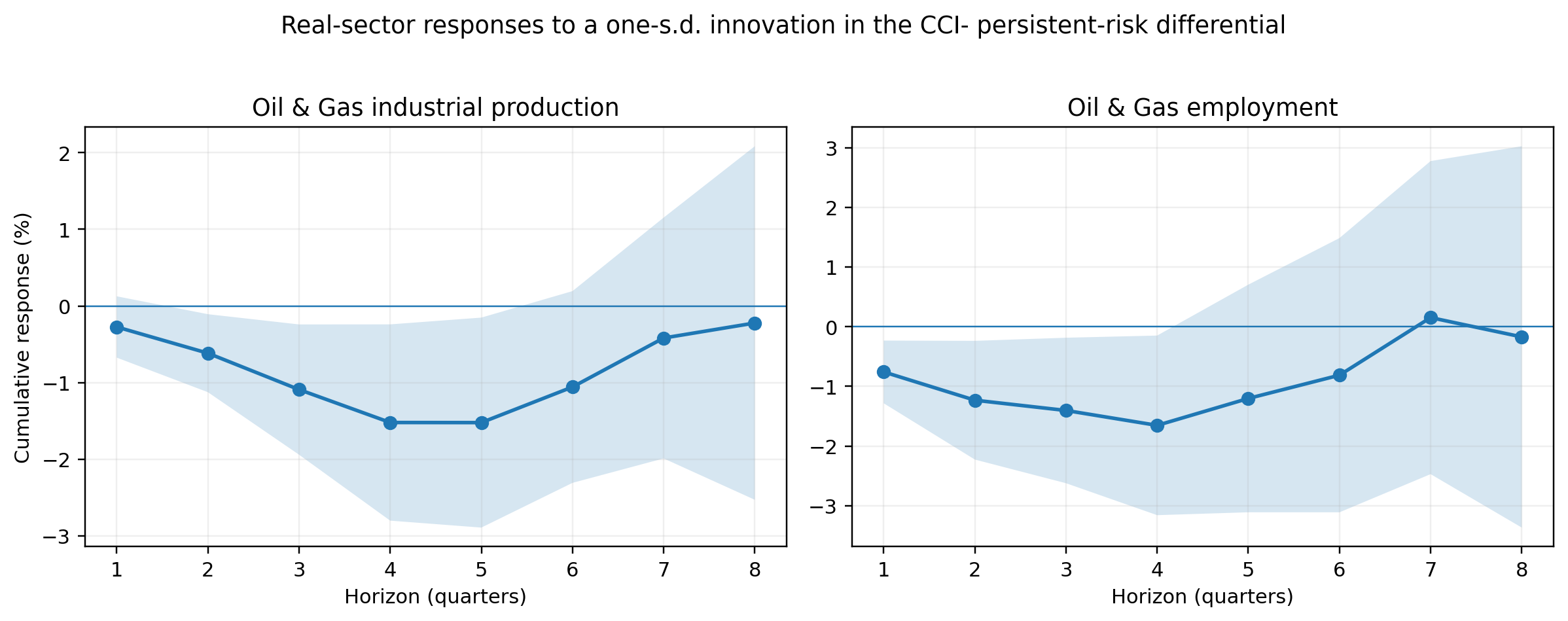}
\caption{Oil and Gas real-sector responses to a one-standard-deviation innovation in the CCI$^-$ persistent-risk differential. The left panel reports cumulative industrial-production responses and the right panel cumulative employment responses. Shaded areas denote 90\% HAC confidence bands.}
\label{fig:realactivity_ip_emp}
\end{figure}

The right panel of Figure~\ref{fig:realactivity_ip_emp} reports the corresponding employment response. The same CCI$^-$ persistent-risk innovation is followed by a decline in Oil and Gas Extraction employment of about $0.8\%$ after one quarter and $1.2\%$ after two quarters. The cumulative response reaches approximately $1.4$--$1.7\%$ over the following two quarters before gradually attenuating. The shorter-horizon responses are estimated relatively precisely, while uncertainty increases at longer horizons. Utilities employment, in contrast, remains close to zero throughout the same horizon.

Taken together, the two panels of Figure~\ref{fig:realactivity_ip_emp} indicate a broadly consistent real-sector response, although the timing differs across outcomes. Employment adjusts earlier, whereas the contraction in industrial production becomes more pronounced after roughly three quarters. Both outcomes therefore point to weaker subsequent activity in the sector for which adverse climate concern is associated with the largest increase in persistent financial risk.

\subsection{Interpretation and robustness}

The production and employment results point to a sector-specific macro-financial relationship. Three features are particularly informative. First, the real-activity responses are concentrated in Oil and Gas rather than Utilities. Second, the clearest production response is associated with the adverse component of climate concern, consistently with the financial evidence in Section~\ref{res}. Third, the predictive content of the persistent-risk innovation is not subsumed by the underlying climate-attention measure itself. When equation~\eqref{eq:lp_real} is augmented with contemporaneous and lagged standardized CCI$^-$, the raw CCI terms are not statistically significant, while the PRD innovation remains negatively associated with subsequent Oil and Gas production at medium horizons. This suggests that the persistent financial-risk transformation contains information about subsequent sectoral activity beyond that in adverse climate attention alone.

A first concern is whether the results depend on the quarterly frequency adopted in the baseline analysis. Appendix~\ref{app:freq} addresses this issue in two ways. Figure~\ref{fig:mixedfreq} retains the quarterly PRD innovation but measures Oil and Gas production responses month by month. The resulting profile closely tracks the quarterly estimates at comparable horizons, with the contraction becoming visible after approximately six months and reaching its largest magnitude between nine and fifteen months. By contrast, Figure~\ref{fig:monthlyprd} shows that constructing and filtering the PRD itself at monthly frequency produces substantially less stable responses. Taken together, these exercises indicate that the baseline results are not generated by quarterly aggregation of the real-activity outcome, while also supporting the use of quarterly innovations for a deliberately slow-moving financial-risk component.

A second concern is the alignment between the financial predictor and the real-sector outcome. The baseline PRD is constructed from the broad Energy and Utilities indices, whereas the main outcomes refer specifically to Oil and Gas Extraction. Appendix~\ref{app:match} therefore reconstructs the CCI$^-$ persistent-risk differential using the US--DS Oil, Gas, Coal equity sub-sector relative to aggregate Utilities. The matched and baseline PRDs are highly correlated, and Figure~\ref{fig:matched_prd_ip} shows that the sector-matched specification generates the same hump-shaped contraction in Oil and Gas industrial production, with somewhat larger point estimates. Employment responses remain negative over the corresponding horizons, although they are less precisely estimated. The matched specification therefore confirms that the production result is not driven by the broader sectoral coverage of the baseline financial predictor.

Additional diagnostics are reported in Appendix~\ref{app:addrob} and summarized in Table~\ref{tab:addrob}. The negative, hump-shaped production response is preserved when the autoregressive filtering of the PRD is modified, when richer lags of sectoral production growth are included, when the Global Financial Crisis or the main pandemic period is excluded, and under alternative HAC bandwidths. The same innovation does not generate a comparable medium-horizon response in aggregate U.S. industrial production or Electric and Gas Utilities production. Leave-one-year-out estimates also retain a negative four-quarter response throughout, although subsample estimates become considerably less precise. Block-bootstrap inference and adjustments across multiple projection horizons are more conservative than the baseline HAC inference.

Overall, these results support a predictive relationship between adverse climate-related persistent financial risk and subsequent weakness in exposed real-sector activity. They do not separately identify the intermediate mechanisms---such as required returns, financing conditions, or investment---and therefore should not be interpreted as establishing a causal transmission channel.

\section{Conclusion}
\label{concl}

This paper studies how public climate concern is reflected in persistent financial risk across U.S. sectors with different exposures to physical and transition risks. Using GARCH--MIDAS models, we combine daily sectoral equity returns with monthly measures of climate concern and separate short-run market volatility from a slow-moving component associated with the CCI.

The financial results reveal a pronounced sectoral asymmetry. Persistent risk in Energy rises with climate concern, particularly with the negative component associated with adverse climate narratives and physical-risk salience. Utilities display a different pattern: broad and transition-oriented climate attention is associated with lower persistent risk at the aggregate level and in several regulated sub-sectors. The more granular evidence shows that the distinction is not simply between ``brown'' Energy and ``green'' Utilities. Pipelines behave differently from upstream Energy activities, while Water exhibits a positive response to climate concern, highlighting the importance of business models and direct exposure to physical climate risk.

The real-sector evidence adds a macro-financial dimension to these findings. Innovations in the persistent-risk differential associated with negative climate concern precede a medium-horizon decline in Oil and Gas industrial production and an earlier decline in Oil and Gas employment, while comparable Utilities responses are limited. The production pattern is also preserved when the financial predictor is reconstructed using the more closely matched Oil, Gas, Coal equity sub-sector. Although statistical precision varies across specifications and inference procedures, the sign and timing of the response are economically stable across a broad set of robustness checks. We therefore interpret these findings as evidence of a predictive link between climate-related persistent financial risk and subsequent activity in exposed sectors, rather than as identification of a causal transmission mechanism.

More broadly, the results indicate that the macroeconomic relevance of climate concern depends importantly on where persistent financial risk is concentrated. Adverse climate attention is not associated with a uniform increase in risk across sectors: its financial and real-sector correlates are strongest in carbon-intensive activities, while regulated infrastructure and physically exposed utility segments display distinct responses. Distinguishing between adverse climate concern and transition-oriented attention, as well as between sectors with different business models and climate exposures, is therefore important for understanding the sectoral consequences of the low-carbon transition. More generally, the findings suggest that macro-financial assessments of climate risk may benefit from explicitly accounting for the heterogeneous way in which climate perceptions are transmitted across financial markets and real-sector activity.

\clearpage

\appendix

\section*{Appendix}

\section{Full GARCH--MIDAS estimates}
\label{app:midas_full}

This appendix reports the complete GARCH--MIDAS estimates underlying the results presented in Section~\ref{res}. The main text focuses on the MIDAS slope coefficient $\theta$, while Tables~\ref{tab:aggregate_midas_full} and~\ref{tab:univariate_6assets_cci_full} report the full parameter estimates for the aggregate sectors and sub-sectors, respectively.

\begin{table}[!htbp]
\centering
\caption{Full GARCH--MIDAS estimates for aggregate Utilities and Energy.}
\label{tab:aggregate_midas_full}
\renewcommand{\arraystretch}{1.15}
\setlength{\tabcolsep}{7pt}

\begin{tabular}{lcccccc}
\toprule
& \multicolumn{3}{c}{US--DS Utilities}
& \multicolumn{3}{c}{US--DS Energy} \\
\cmidrule(lr){2-4}
\cmidrule(lr){5-7}
Parameter
& Full & CCI$^-$ & CCI$^+$
& Full & CCI$^-$ & CCI$^+$ \\
\midrule

$\mu$
& \estse{-0.0208}{0.0102}
& \estse{-0.0218}{0.0100}
& \estse{-0.0205}{0.0102}
& \estse{-0.0438}{0.0113}
& \estse{-0.0453}{0.0114}
& \estse{-0.0439}{0.0113} \\

$\alpha$
& \estse{0.0583}{0.0043}
& \estse{0.0538}{0.0037}
& \estse{0.0591}{0.0045}
& \estse{0.0622}{0.0020}
& \estse{0.0643}{0.0022}
& \estse{0.0615}{0.0020} \\

$\beta$
& \estse{0.9281}{0.0056}
& \estse{0.9391}{0.0043}
& \estse{0.9225}{0.0064}
& \estse{0.9335}{0.0026}
& \estse{0.9293}{0.0030}
& \estse{0.9346}{0.0025} \\

$\theta$
& \estse{-6.5077}{1.3172}
& \estse{0.5803}{0.3700}
& \estse{-5.6869}{0.8704}
& \estse{1.3343}{0.3788}
& \estse{4.8184}{1.6667}
& \estse{1.1141}{0.3763} \\

$\omega$
& \estse{1.0010}{0.0765}
& \estse{71.1220}{105.6000}
& \estse{1.0010}{0.0803}
& \estse{76.9870}{58.0370}
& \estse{4.5201}{2.1911}
& \estse{79.3620}{73.6490} \\

$\kappa$
& \estse{3.4340}{0.7856}
& \estse{-0.6755}{0.2470}
& \estse{2.9805}{0.5309}
& \estse{-0.5501}{0.3535}
& \estse{-2.4664}{0.9397}
& \estse{-0.4069}{0.3650} \\

\bottomrule
\end{tabular}

\begin{minipage}{0.92\textwidth}
\footnotesize
\emph{Notes:} The table reports the complete maximum-likelihood estimates
of the GARCH--MIDAS specifications underlying
Table~\ref{tab:aggregate_midas}.
Standard errors are reported in parentheses.
\end{minipage}
\end{table}

\begin{table}[!htbp]
\centering
\caption{Full GARCH--MIDAS estimates by sub-sector and CCI specification.}
\label{tab:univariate_6assets_cci_full}
\renewcommand{\arraystretch}{1.15}
\setlength{\tabcolsep}{5pt}
\scriptsize

\resizebox{\textwidth}{!}{%
\begin{tabular}{l*{9}{c}}
\toprule
& \multicolumn{3}{c}{US--DS Oil, Gas, Coal}
& \multicolumn{3}{c}{US--DS Integrated Oil and Gas}
& \multicolumn{3}{c}{US--DS Pipelines} \\
\cmidrule(lr){2-4}
\cmidrule(lr){5-7}
\cmidrule(lr){8-10}

Parameter
& Full & CCI$^-$ & CCI$^+$
& Full & CCI$^-$ & CCI$^+$
& Full & CCI$^-$ & CCI$^+$ \\
\midrule

$\mu$
& \estse{-0.0445}{0.0117}
& \estse{-0.0459}{0.0118}
& \estse{-0.0465}{0.0120}
& \estse{-0.0434}{0.0113}
& \estse{-0.0442}{0.0113}
& \estse{-0.0436}{0.0113}
& \estse{0.0037}{0.0122}
& \estse{0.0040}{0.0123}
& \estse{0.0036}{0.0122} \\

$\alpha$
& \estse{0.0651}{0.0021}
& \estse{0.0681}{0.0023}
& \estse{0.0670}{0.0030}
& \estse{0.0644}{0.0041}
& \estse{0.0662}{0.0043}
& \estse{0.0634}{0.0039}
& \estse{0.1236}{0.0037}
& \estse{0.1227}{0.0035}
& \estse{0.1235}{0.0037} \\

$\beta$
& \estse{0.9301}{0.0029}
& \estse{0.9248}{0.0033}
& \estse{0.9275}{0.0034}
& \estse{0.9301}{0.0045}
& \estse{0.9260}{0.0049}
& \estse{0.9315}{0.0042}
& \estse{0.8553}{0.0056}
& \estse{0.8575}{0.0055}
& \estse{0.8538}{0.0057} \\

$\theta$
& \estse{1.1455}{0.3513}
& \estse{4.3827}{1.7203}
& \estse{-6.6716}{1.6034}
& \estse{1.1326}{0.3232}
& \estse{4.9426}{1.5948}
& \estse{0.8084}{0.3402}
& \estse{-3.5002}{1.1391}
& \estse{-2.4235}{1.4415}
& \estse{-3.1972}{0.8459} \\

$\omega$
& \estse{113.0300}{144.2300}
& \estse{4.2784}{2.3149}
& \estse{1.0010}{0.0816}
& \estse{581.7000}{3316300.0000}
& \estse{4.1454}{2.0134}
& \estse{546.3300}{2334400.0000}
& \estse{1.0010}{0.1483}
& \estse{1.0010}{0.2059}
& \estse{1.0010}{0.1561} \\

$\kappa$
& \estse{-0.3648}{0.3365}
& \estse{-2.1468}{0.9652}
& \estse{4.3421}{1.0395}
& \estse{-0.4898}{0.2983}
& \estse{-2.5931}{0.8854}
& \estse{-0.2838}{0.3178}
& \estse{2.3966}{0.6929}
& \estse{1.6452}{0.7921}
& \estse{2.2286}{0.5272} \\

\midrule

& \multicolumn{3}{c}{US--DS Gas, Water, Multi-Utilities}
& \multicolumn{3}{c}{US--DS Gas Distribution}
& \multicolumn{3}{c}{US--DS Water} \\
\cmidrule(lr){2-4}
\cmidrule(lr){5-7}
\cmidrule(lr){8-10}

Parameter
& Full & CCI$^-$ & CCI$^+$
& Full & CCI$^-$ & CCI$^+$
& Full & CCI$^-$ & CCI$^+$ \\
\midrule

$\mu$
& \estse{-0.0158}{0.0097}
& \estse{-0.0162}{0.0097}
& \estse{-0.0155}{0.0097}
& \estse{-0.0139}{0.0103}
& \estse{-0.0142}{0.0103}
& \estse{-0.0143}{0.0100}
& \estse{-0.0066}{0.0138}
& \estse{-0.0064}{0.0138}
& \estse{-0.0069}{0.0138} \\

$\alpha$
& \estse{0.0662}{0.0047}
& \estse{0.0630}{0.0042}
& \estse{0.0679}{0.0050}
& \estse{0.0658}{0.0045}
& \estse{0.0638}{0.0042}
& \estse{0.0789}{0.0053}
& \estse{0.0890}{0.0062}
& \estse{0.0894}{0.0064}
& \estse{0.0896}{0.0061} \\

$\beta$
& \estse{0.9210}{0.0058}
& \estse{0.9293}{0.0048}
& \estse{0.9140}{0.0066}
& \estse{0.9183}{0.0057}
& \estse{0.9248}{0.0051}
& \estse{0.9063}{0.0062}
& \estse{0.8680}{0.0101}
& \estse{0.8636}{0.0107}
& \estse{0.8696}{0.0099} \\

$\theta$
& \estse{-7.9874}{1.5835}
& \estse{-7.5595}{2.3491}
& \estse{-7.0016}{1.0357}
& \estse{-5.6926}{1.3218}
& \estse{-5.0599}{1.8743}
& \estse{-4.7346}{1.1221}
& \estse{3.1749}{0.5017}
& \estse{4.5796}{0.6743}
& \estse{2.0402}{0.3851} \\

$\omega$
& \estse{1.0010}{0.0689}
& \estse{1.0010}{0.0653}
& \estse{1.0010}{0.0758}
& \estse{1.0010}{0.0760}
& \estse{1.0010}{0.0839}
& \estse{4.2462}{2.1924}
& \estse{7.8838}{2.8848}
& \estse{3.6202}{1.0717}
& \estse{13.5020}{5.8031} \\

$\kappa$
& \estse{4.3087}{0.9434}
& \estse{3.7013}{1.2825}
& \estse{3.7699}{0.6300}
& \estse{3.0642}{0.7835}
& \estse{2.4636}{1.0153}
& \estse{2.5759}{0.6870}
& \estse{-1.6865}{0.3053}
& \estse{-2.2779}{0.3707}
& \estse{-1.0264}{0.2426} \\

\bottomrule
\end{tabular}%
}

\begin{minipage}{0.95\textwidth}
\footnotesize
\emph{Notes:} The table reports the complete maximum-likelihood estimates
of the GARCH--MIDAS specifications underlying
Table~\ref{tab:univariate_6assets_cci}.
Standard errors are reported in parentheses.
\end{minipage}
\end{table}

\section{PRD innovation diagnostics}
\label{app:prd}

The persistent-risk differentials used in the local projections are highly
persistent. As described in Section~\ref{macro}, we therefore extract
innovations from autoregressive models before estimating the real-activity
responses. Table~\ref{tab:prd_ar_diagnostics} reports the selected lag orders
and residual-autocorrelation diagnostics.

\begin{table}[!htbp]
\centering
\caption{Autoregressive specification and residual diagnostics for the persistent-risk differentials.}
\label{tab:prd_ar_diagnostics}
\renewcommand{\arraystretch}{1.15}
\setlength{\tabcolsep}{10pt}

\begin{tabular}{lccccc}
\toprule
PRD specification
& Selected AR order
& AIC
& BIC
& LB(4)
& LB(8) \\
\midrule

Full CCI
& 5
& -109.86
& -94.64
& 0.294
& 0.650 \\

CCI$^-$
& 6
& -166.20
& -148.93
& 0.967
& 0.817 \\

CCI$^+$
& 5
& -141.48
& -126.26
& 0.334
& 0.759 \\

\bottomrule
\end{tabular}

\begin{minipage}{0.90\textwidth}
\footnotesize
\emph{Notes:} Candidate models are AR($p$) specifications with
$p=1,\ldots,8$. AIC and BIC denote the Akaike and Bayesian information
criteria for the selected model. LB(4) and LB(8) report $p$-values from
Ljung--Box tests of residual serial correlation at lags 4 and 8,
respectively. Residuals from the selected models are standardized to unit
variance before entering the local projections.
\end{minipage}
\end{table}

\section{Frequency robustness of the real-activity exercise}
\label{app:freq}

The baseline local projections are conducted at quarterly frequency. Since
both the underlying GARCH--MIDAS persistent component and the Federal
Reserve industrial-production series are available monthly, we examine
whether the baseline findings are sensitive to temporal aggregation.

We first retain the quarterly PRD innovation used in the baseline
specification but measure the Oil and Gas production response at monthly
horizons. Each observation is dated at the final month of the quarter over
which the PRD is constructed, and the dependent variable is the cumulative
change in monthly Oil and Gas industrial production from that month to
horizon $h=1,\ldots,24$. The controls follow the quarterly baseline, while
the HAC bandwidth is expressed in quarterly observations and increases with
the projection horizon. This exercise changes the frequency at which the
real response is observed without redefining the persistent financial-risk
innovation.

\begin{figure}[!htbp]
\centering
\includegraphics[width=.78\textwidth]{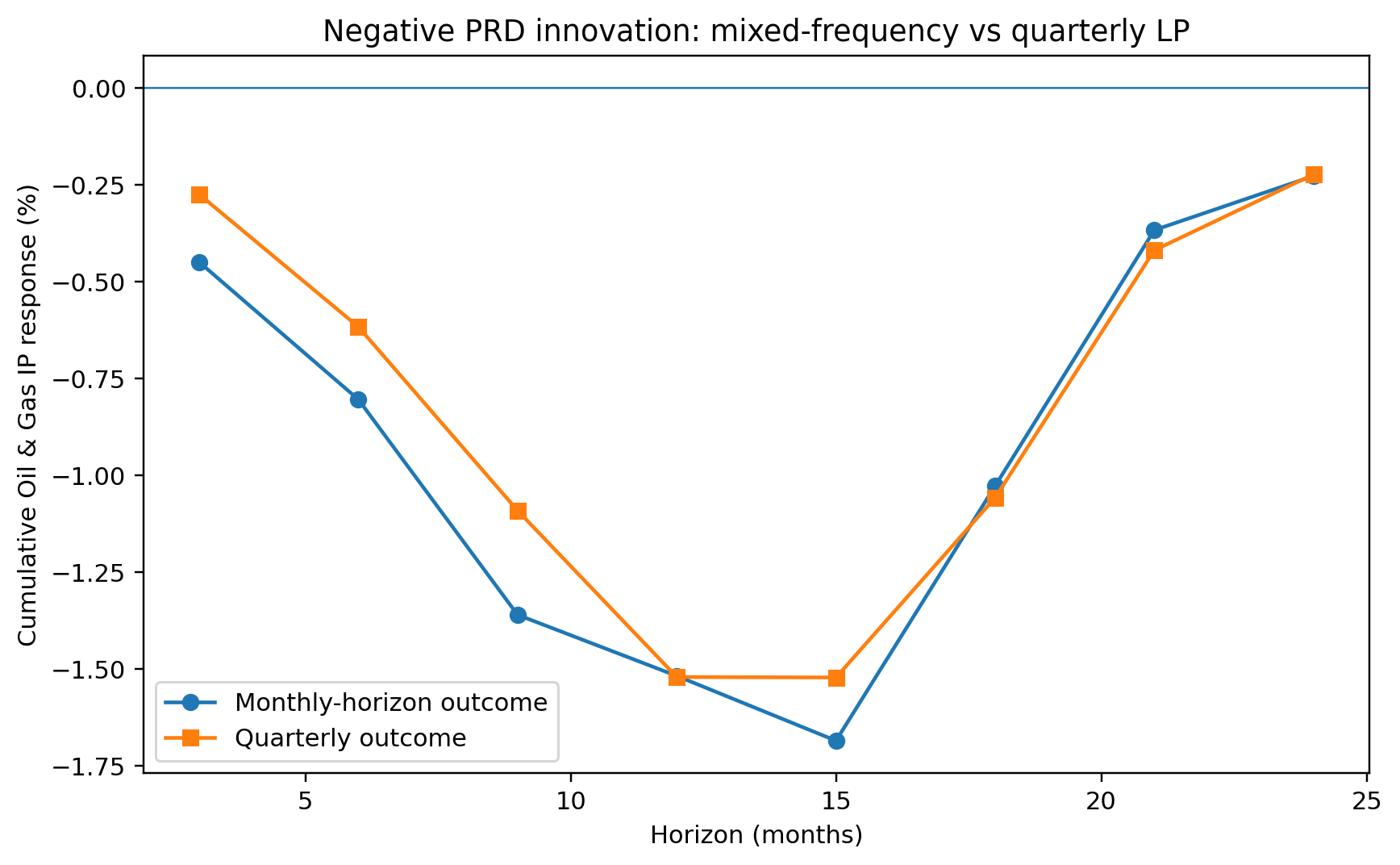}
\caption{Oil and Gas industrial-production responses to a quarterly
innovation in the CCI$^-$ persistent-risk differential. The
monthly-horizon specification measures the cumulative response using
monthly industrial production from the end of each quarter, while the
quarterly specification uses the baseline quarterly outcome.}
\label{fig:mixedfreq}
\end{figure}

Figure~\ref{fig:mixedfreq} compares the monthly-resolved response with the
quarterly baseline at common horizons. The two profiles are closely aligned.
The monthly-resolved response is approximately $-0.8\%$ after six months,
$-1.4\%$ after nine months, $-1.5\%$ after twelve months, and $-1.7\%$
after fifteen months. The nine-month response is estimated relatively
precisely, while uncertainty increases at the twelve- and fifteen-month
horizons. The similarity of the two profiles indicates that the
medium-horizon contraction is not mechanically generated by quarterly
averaging of industrial production.

\begin{figure}[!htbp]
\centering
\includegraphics[width=.78\textwidth]{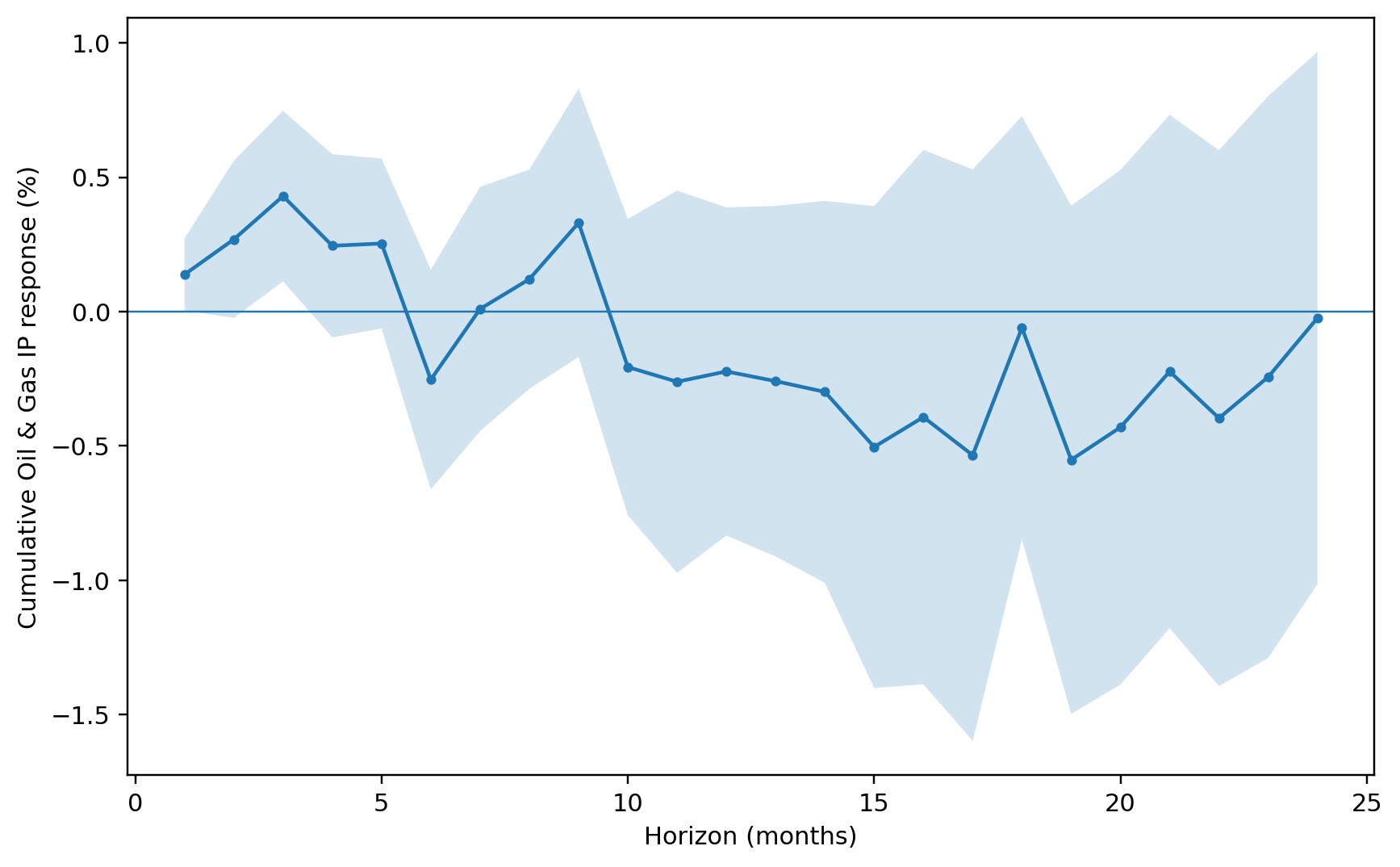}
\caption{Cumulative Oil and Gas industrial-production response when both
the CCI$^-$ persistent-risk differential innovation and industrial
production are defined at monthly frequency. Shaded areas denote 90\% HAC
confidence bands.}
\label{fig:monthlyprd}
\end{figure}

As a second check, we construct the PRD and extract its innovation directly
at monthly frequency. Figure~\ref{fig:monthlyprd} shows that this alternative
produces substantially less stable responses: a small positive response
appears at short horizons, while the medium-horizon negative response is
modest and imprecisely estimated. The contrast suggests that the results are
more sensitive to the frequency at which innovations are extracted from the
slow-moving PRD than to the frequency at which real activity is measured.
This is consistent with using quarterly innovations as the baseline for a
financial component designed to capture persistent rather than
high-frequency risk.

\section{Sector matching and employment}
\label{app:match}

The baseline PRD is constructed from the broad Energy and Utilities indices,
while the main real-sector outcomes refer specifically to Oil and Gas
Extraction. We therefore construct an alternative financial predictor using
the US--DS Oil, Gas, Coal sub-sector. Specifically, we replace the aggregate
Energy persistent component with the CCI$^-$ GARCH--MIDAS component
estimated for Oil, Gas, Coal in Section~\ref{sub}, while retaining aggregate
Utilities as the comparison sector. The resulting matched PRD is highly
correlated with the broad Energy--Utilities PRD, with correlations of
approximately 0.99 in levels and 0.96 in first differences.

\begin{figure}[!htbp]
\centering
\includegraphics[width=.82\textwidth]{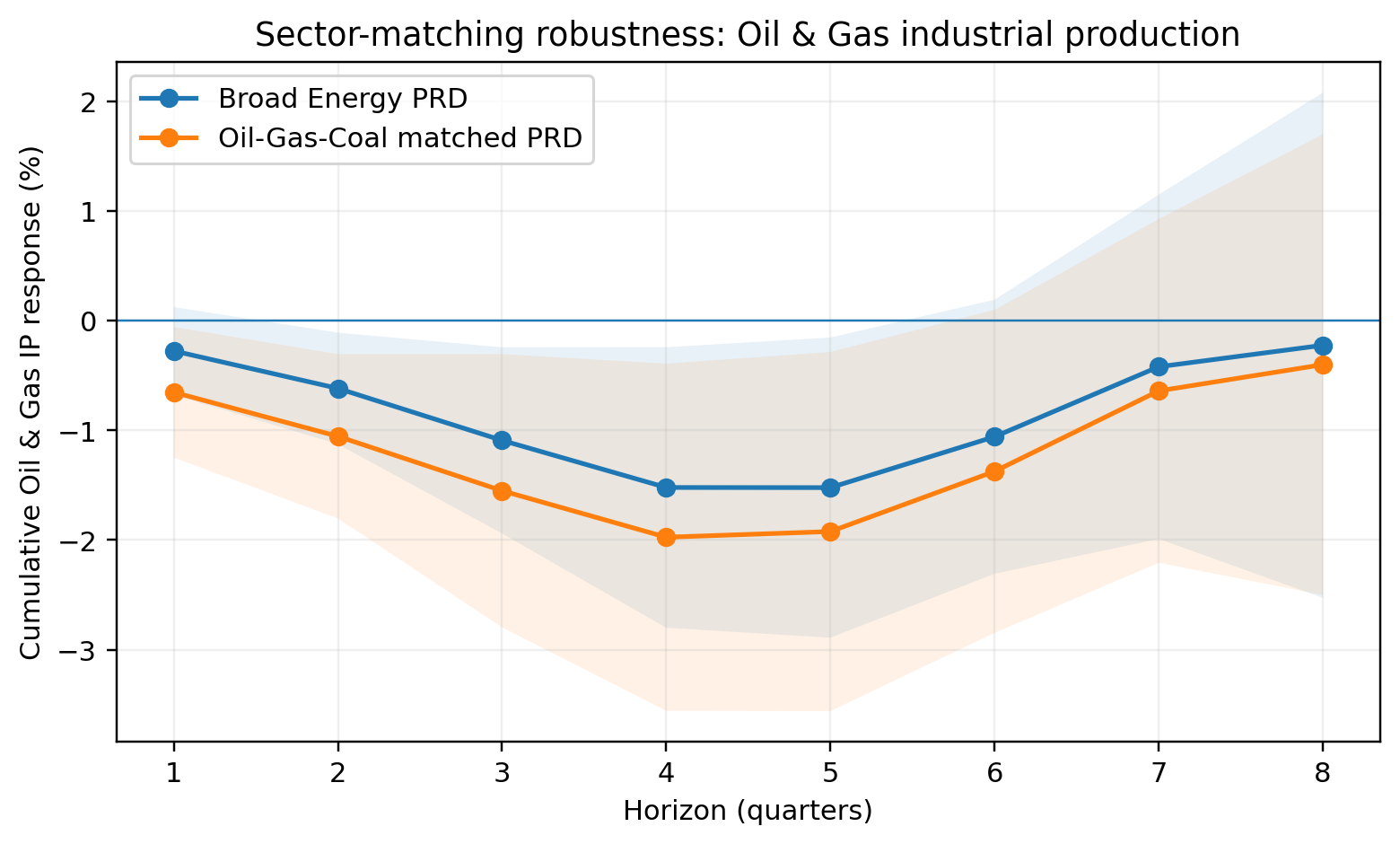}
\caption{Sector-matching robustness for Oil and Gas industrial production.
The broad specification uses the aggregate Energy--Utilities CCI$^-$
persistent-risk differential; the matched specification replaces broad
Energy with the US--DS Oil, Gas, Coal sub-sector. Shaded areas denote 90\%
HAC confidence bands.}
\label{fig:matched_prd_ip}
\end{figure}

Figure~\ref{fig:matched_prd_ip} compares the Oil and Gas
industrial-production responses obtained from the broad and sector-matched
PRD innovations. The matched specification produces the same hump-shaped
contraction and somewhat larger point estimates: approximately $-1.05\%$
after two quarters, $-1.55\%$ after three, and $-1.97\%$ after four
quarters. The corresponding $p$-values are 0.025, 0.046, and 0.046,
respectively. The sector-matching exercise therefore indicates that the
production result is not driven by the broader sectoral coverage of the
baseline financial predictor.

For Oil and Gas employment, the matched CCI$^-$ PRD produces negative point
estimates over the first four quarters, of approximately $-0.5\%$,
$-0.7\%$, $-0.8\%$, and $-1.1\%$, respectively. The confidence intervals
are wider than for the corresponding production responses. Thus, the
sector-matching exercise preserves the direction of the employment response,
although less precisely than for industrial production.

\section{Additional robustness and diagnostic checks}
\label{app:addrob}

This section reports additional diagnostics for the predictive
interpretation of the local-projection evidence. We consider whether the
PRD innovation reflects direct predictive content in the underlying CCI,
whether it responds systematically to prior Oil and Gas activity, and how
the results change under alternative dynamic specifications and inference
procedures.

We first augment the baseline local projection with contemporaneous and
lagged standardized CCI$^-$. The raw CCI terms are not statistically
significant at the horizons at which the Oil and Gas response is most
pronounced, while the coefficient on the PRD innovation remains negative
and informative. At horizons three and four, the cumulative responses to
the PRD innovation are approximately $-1.58\%$ and $-1.84\%$, with
$p$-values of 0.020 and 0.027, respectively. Thus, the predictive content
of the persistent-risk innovation is not subsumed by the level of adverse
climate attention itself.

We next examine whether the quarterly CCI$^-$ PRD innovation is predictable
from prior sectoral activity. Regressing the standardized innovation on
four lags of Oil and Gas production growth, together with lagged oil-price,
aggregate-activity, and monetary-policy controls, does not reject the joint
null that the four sectoral growth lags have zero coefficients
($p=0.268$). A two-quarter cumulative pre-movement is more informative and
is borderline predictive in an alternative specification, however. We
therefore do not interpret the PRD innovation as an externally identified
structural shock.

Consistent with this qualification, enriching the local projection with
four additional lags of Oil and Gas production growth leaves the
medium-horizon point estimates economically similar but less precise. The
responses at horizons three, four, and five are approximately $-1.15\%$,
$-1.54\%$, and $-1.80\%$, with $p$-values of 0.067, 0.103, and 0.094,
respectively.

The production response is also not tied to the exact autoregressive
filtering of the PRD. When we first difference the PRD and extract an
autoregressive innovation from $\Delta PRD$, the response remains negative
and hump shaped, with a four-quarter response of approximately $-1.07\%$
($p=0.079$). Moreover, the pattern is sector specific: the same CCI$^-$
PRD innovation does not generate a comparable medium-horizon response in
aggregate U.S. industrial production or Electric and Gas Utilities
production. At four quarters, the corresponding coefficients are
approximately $-0.46\%$ ($p=0.456$) and $-0.10\%$ ($p=0.859$),
respectively.

Stability checks lead to a similar conclusion. Re-estimating the
four-quarter projection after dropping one calendar year at a time produces
a negative coefficient in every case, ranging from approximately $-0.43\%$
to $-2.48\%$. A simple pre/post-2017 split yields a larger response in the
earlier subsample and a smaller, imprecisely estimated response after 2017.
Given the limited number of usable observations in each subsample, we view
this exercise as a stability diagnostic rather than evidence of a
structural break.

Finally, inference becomes more conservative under procedures that account
more aggressively for sampling uncertainty across horizons. Block-bootstrap
bands are wider than the baseline asymptotic HAC intervals, while Holm and
false-discovery-rate adjustments across the eight projection horizons
remove conventional pointwise significance. These checks affect statistical
precision more than the sign and timing of the estimated response.

Table~\ref{tab:addrob} summarizes the main robustness results.

\begin{table}[!htbp]
\centering
\caption{Selected robustness checks for the CCI$^-$ persistent-risk differential.}
\label{tab:addrob}
\renewcommand{\arraystretch}{1.10}
\setlength{\tabcolsep}{7pt}

\begin{tabular}{lccc}
\toprule
Specification / outcome & Horizon & Response (\%) & $p$-value \\
\midrule

Baseline Oil and Gas IP
& 3 qtrs & -1.09 & 0.040 \\

Baseline Oil and Gas IP
& 4 qtrs & -1.52 & 0.056 \\

PRD controlling for raw CCI$^-$
& 3 qtrs & -1.58 & 0.020 \\

PRD controlling for raw CCI$^-$
& 4 qtrs & -1.84 & 0.027 \\

Current + 4 lags of Oil and Gas IP growth
& 3 qtrs & -1.15 & 0.067 \\

Current + 4 lags of Oil and Gas IP growth
& 4 qtrs & -1.54 & 0.103 \\

Alternative innovation in $\Delta PRD$
& 4 qtrs & -1.07 & 0.079 \\

Aggregate U.S. industrial production
& 4 qtrs & -0.46 & 0.456 \\

Electric and Gas Utilities production
& 4 qtrs & -0.10 & 0.859 \\

Matched Oil, Gas, Coal PRD: Oil and Gas IP
& 4 qtrs & -1.97 & 0.046 \\

Baseline Oil and Gas employment
& 1 qtr & -0.76 & 0.021 \\

Baseline Oil and Gas employment
& 2 qtrs & -1.23 & 0.047 \\

\bottomrule
\end{tabular}

\begin{minipage}{0.94\textwidth}
\footnotesize
\emph{Notes:} Responses are cumulative log changes following a
one-standard-deviation innovation. The baseline uses the specification in
equation~\eqref{eq:lp_real}. The specification controlling for raw CCI$^-$
adds contemporaneous and lagged standardized adverse climate concern. The
richer dynamic specification adds four lags of sectoral
industrial-production growth.
\end{minipage}
\end{table}

\clearpage

\bibliographystyle{plainnat}
\bibliography{references}

\end{document}